\documentclass[aps,pre,onecolumn,10pt]{revtex4}
\usepackage{amsfonts,amscd}
\usepackage{amssymb}
\usepackage{mathtools}
\usepackage{amsmath}
\usepackage{amscd}
\usepackage{amsmath}
\usepackage{lineno,hyperref}
\modulolinenumbers[5]
\usepackage{float}
\usepackage{xcolor}
\usepackage[]{graphicx}
\usepackage{subfigure} 
\usepackage{chngcntr}       
\begin{document}        
\title{Chimera states formed via a two-level synchronization mechanism}     
\author{A. Provata}                           
\affiliation{Institute of Nanoscience and Nanotechnology,     
NCSR ``Demokritos'', GR-15341 Athens, Greece                     
}     
\date{today}             

\begin{abstract}
Chimera states, which consist of coexisting synchronous and asynchronous domains
in networks of coupled oscillators, are in the focus of attention for over a
decade. Although chimera morphology and properties have been investigated
in a number of models, the mechanism responsible for their formation 
is still not well understood.
To shed light in the chimera producing mechanism,
in the present study we introduce an oscillatory model with variable frequency
governed by a 3rd order equation. In this model single oscillators are
constructed as bistable and depending
on the initial conditions their frequency may result in one of the two 
stable fixed points, $\omega _l$ and $\omega _h $ (two-level synchronization).
Numerical simulations demonstrate that these oscillators 
organize in domains with alternating frequencies, when they
are nonlocally coupled in networks.  In each domain the oscillators synchronize,
sequential domains follow different modes of synchronization and the border elements
between two consecutive domains form the asynchronous domains as they are 
influenced by both frequencies.  We investigate the influence of the frequency coupling constant
and of the coupling range on the chimera morphology and we show that the chimera
multiplicity decreases as the coupling range increases. For small values of the frequency 
coupling constant two coherent (incoherent) domains are formed, for intermediate values
the number of domains increases, while for larger values some frequency domains are
absorbed by others and synchronization settles. The frequency spectrum is calculated in
the coherent and incoherent domains of this model. In the coherent domains single
frequencies  ($\omega _l$ or $\omega _h$) are observed, 
while in the incoherent domains both  $\omega _l$ and $\omega _h$ as well as their
superpositions appear. This mechanism of creating domains of
alternating frequencies offers a reasonable generic scenario for chimera state formation.

\end{abstract}
\maketitle

% \vskip 0.7cm            

%--------------------------------------------------------------------------------
\everymath{\displaystyle}

\section{Introduction}  
\label{sec:intro}
%Motivation
\par Systems of nonlocally coupled oscillatory elements often split in domains where the elements
oscillate coherently and other domains where the oscillators are incoherent. These counterintuitive, hybrid
states are known in the literature as ``chimera states'' and they occur even if
 the oscillators are identical and identically linked 
\cite{panaggio:2015,schoell:2016,zheng:2016,oomelchenko:2018}. Chimera states were 
first reported in 2002 by Kuramoto and Battogtokh \cite{kuramoto:2002, kuramoto:2002a},
while the term ``chimera state'' was proposed two years later by Abrams and Strogatz \cite{abrams:2004}.
In the original articles the Kuramoto phase oscillator was  used and, later on, chimera states
were confirmed in diverse oscillatory flows such as the FitzHugh-Nagumo, the Van der Pol,
the Stuart-Landau, the Hindmarsh-Rose and Integrate-and-Fire models 
\cite{omelchenko:2013,ulonska:2016,tumash:2019,gjurchinovski:2017,zakharova:2017,hizanidis:2014,olmi:2015,goldschmidt:2019,tsigkri:2016}.
After the pioneer studies in oscillatory flows, chimeras are now frequently observed in coupled chaotic
oscillators \cite{shepelev:2017,bukh:2017} and in coupled discreet maps \cite{malchow:2018} 
under a variety of coupling schemes and
parameter values.  
\par Experimentally, chimera states have been reported in 
coupled mechanical oscillators \cite{martens:2013,blaha:2016,dudkowski:2016},
chemical oscillators \cite{tinsley:2012,nkomo:2013,taylor:2015},
electronic oscillators \cite{rosin:2014,english:2017},
 nonlinear optics \cite{hagerstrom:2012} and in laser physics \cite{uy:2019}. 
In nature, chimeras have been associated with the unihemispheric sleep of birds and mammals 
\cite{rattenborg:2000,rattenborg:2006} and with brain malfunctions such as the onset of seizures in epilepsy
  \cite{mormann:2000,mormann:2003a,andrzejak:2016}.
\par Despite the extensive numerical evidence of chimera states, their confirmation in many dynamical
systems and the experimental observations,
 the mechanism behind the formation of chimeras remains elusive. Earlier studies have focused on 
analytical approaches to the Kuramoto model \cite{kuramoto:2002,abrams:2004,panaggio:2015,oomelchenko:2018} 
while most recent approaches embrace the idea of the presence of bistable elements in the system
\cite{hizanidis:2014,dudkowski:2014,shepelev:2017,shepelev:2019}.
Along these lines, in the present study we propose a toy model to explore further the idea of chimera states
produced as a result of the presence of bistable elements. To pursue this idea,  
an oscillatory toy model
of minimal complexity is constructed: it consists of a circularly limiting orbit of constant radius, it has
 constant mean phase velocity
around the orbit, while the dynamical approach to this orbit is of purely exponential type. The 
uncoupled toy model has an explicit analytical solution \cite{provata:2018}. When many toy-oscillators
are coupled in a network, the numerically integrated system does not lead to chimera states.
We show numerically that  it is possible
to achieve chimera states by allowing the toy-oscillators to choose between two different frequency levels.
This type of chimera states presents, by construction,
 two levels of synchronization while the dynamics together with the coupling cause the formation 
of alternating domains with different
frequency which are mediated by the incoherent domains. These incoherent domains
 serve as transition regions and are characterized
by a gradient in the values of the mean phase velocities. The proposed mechanism is generic and can
be the cause behind many of the systems exhibiting known chimera states. 

%Organization
\par The organization of the work is as follows: 
In sec.~\ref{sec:model} we introduce the uncoupled two-frequency nonlinear toy-oscillator
and we discuss its steady state properties.
In sec.~\ref{sec:coupled} we couple the toy-oscillators in a ring network and we discuss
synchronization measures. In sec.~\ref{sec:range} we examine the emerging chimera morphologies and their
multiplicity as a function of the coupling range. In sec.~\ref{sec:freq} we vary the 
frequency coupling which influences the formation of frequency domains and we record the
corresponding variations in the chimera states. In sec.\ref{sec:spectra}
we analyze the frequency spectra of the oscillators belonging to the coherent and incoherent
domains and we show that the incoherent oscillations are the result of the superposition of
the two bordering frequencies.
In the Concluding section we recapitulate our main results and discuss open problems.

\section{The two-frequency oscillator}
\label{sec:model}
The motivation for introducing the present model is the need of a nonlinear oscillator
with explicitly controllable frequency. In this respect we have introduce in Ref.~\cite{provata:2018}
a model oscillator whose trajectory tends exponentially to a closed limiting orbit. We will make
use of this model which, throughout its trajectory, keeps a constant, explicit  frequency $\omega$,
externally controllable as desired. In the following we use the terms ``frequency'', or ``mean phase velocity''
or ``angular frequency'', interchangeably,  to refer to the values of $\omega$, although the angular frequency
is related to the frequency $f$ by a factor of $2\pi$, $\omega =2\pi f$.
 The model, before frequency modulation, 
has the following form \cite{provata:2018}:
\begin{subequations}
\begin{align}
\frac{dx}{dt} &= -ax+\frac{aR}{\sqrt{x^2+y^2}}x-\omega y\\
\frac{dy}{dt} &= -ay+\frac{aR}{\sqrt{x^2+y^2}}y+ \omega x
\end{align}
\label{eqno01}
\end{subequations}

This system presents exponential relaxation to a circle with radius $R$, with relaxation exponent $a$. The solution 
of Eq.~\ref{eqno01} can be explicitly written:

\begin{subequations}
\begin{align}
x(t) &= R(1-Ae^{-at}) \cos (\omega t)\\
y(t) &= R(1-Ae^{-at}) \sin (\omega t)
\end{align}
\label{eqno02}
\end{subequations}
In Eq.~\ref{eqno02} the system starts from position $\left( x_0,y_0\right) : \> x_0^2+y_0^2=R^2(1-A)^2$, 
where $A$ determines the
initial condition inside a circle of radius $R$. As time increases the term $Ae^{-at}$ decreases exponentially to $0$, giving rise to a purely circular orbit. 
It is possible to use a similar construction for the case on a limiting orbit with non-equal
axes, $R_1$ and $R_2$ (an ellipse) \cite{provata:2018}. Without loss of generality
we will study here the case $R_1=R_2=R$, to keep the computations as simple as possible and with
a minimum number of parameters.
\par
\begin{figure}[!h]
\includegraphics[clip,width=0.55\linewidth,angle=0]{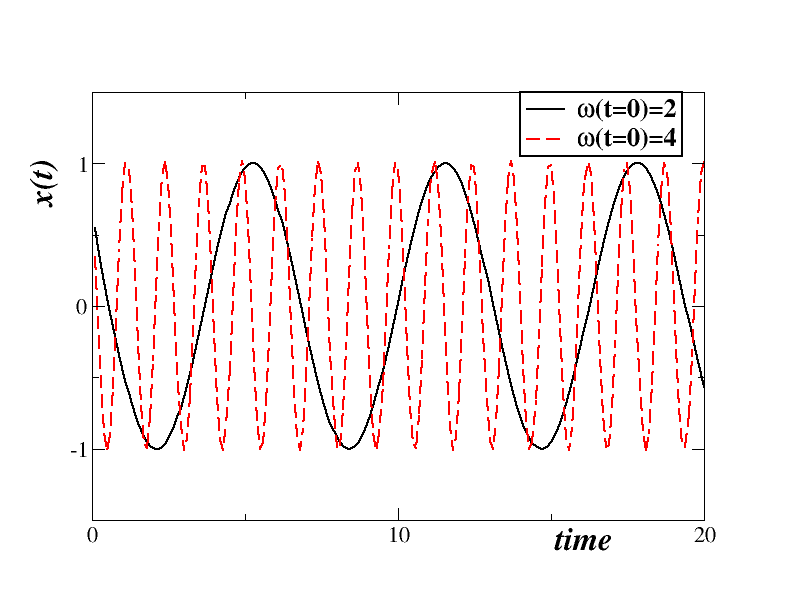}
\caption{\label{fig01} (Color online) Uncoupled oscillators:
The time evolution of the $x-$ variable for $\omega(t=0)=2$ (black solid line) and for $\omega(t=0)=4.0$ (red, dashed line).
All other parameters are common and are set to: $c=-1$, $\omega_l=1$, $\omega_c=3$, $\omega_h=5$, $R=1$ and $a=1.0$. 
}
\end{figure}

To modulate the frequency $\omega$ we introduce a third equation treating $\omega$ as variable. The resulting 
toy-oscillator then takes the form:
\begin{subequations}
\begin{align}
\frac{dx}{dt} &= -ax+\frac{aR}{\sqrt{x^2+y^2}}x-\omega y
\label{eqno03a} \\
\frac{dy}{dt} &= -ay+\frac{aR}{\sqrt{x^2+y^2}}y+ \omega x 
\label{eqno03b} \\
\frac{d\omega}{dt} &= c (\omega -\omega_l)(\omega -\omega_c)(\omega -\omega_h)
\label{eqno03c}
\end{align}
\label{eqno03}
\end{subequations}
In Eq.~\ref{eqno03c} $c$ is a constant, and $\omega_l, \> \omega_c, \>
\omega_h$ (standing for low-$\omega$, intermediate-$\omega$ and high-$\omega$) represent the three fixed points of Eq.~\ref{eqno03c}. Depending on the value of $c$ we can have: 
\begin{enumerate}
\item  one attracting fixed point ($\omega_c$), and two repulsive ones ($\omega_l, \> \omega_h$) if $c>0$ or
\item  two attracting fixed points ($\omega_l, \>\omega_h$), and one repulsive one ($\omega_c$) if $c<0$. 
\end{enumerate}
We are interesting in the second case of bistability  (2), where depending on the initial frequencies the system can fall in either one
of the two attracting fixed points. 
An example is given in Fig.~\ref{fig01}, where starting at $t=0$ from $ \omega(0)=2$ we end-up
having frequency $\omega_l=1$ (black, solid line), while starting from $\omega(0)=4$ we end-up
having frequency $\omega_h=5$ (red, dashed line). In the next sections, when many oscillators starting from random initial conditions $\left( x(t=0),y(t=0),\omega(t=0)\right)$
will be coupled in the network, a number of them will tend to the $\omega_l$ fixed point while the rest
will end up on the $\omega_h$ fixed point depending on their initial condition,
see sec.~\ref{sec:coupled}.
%\par Without loss of generality, in the following sections
%we will use as working parameters $c=-1$ to ensure of the existence of one repulsive fixed point ($\omega_c=3$)
%surrounded by two attractive ones ($\omega_l=1$, $\omega_h=5$), $R=1$ and $a=1.0$.

\section{Two-frequency oscillators coupled in a ring network}
\label{sec:coupled}

In this section we couple the two-frequency oscillators in a ring network arrangement. In a network containing
$N$ elements, we use the simplest nonlocal coupling scheme 
where each oscillator is coupled linearly to $2S$ neighbors: $S$ nearest neighbors on its left and $S$ on its right. Moreover, 
 each $x_i-$variable is only coupled to $x_j$-variables, $j=i-S, \cdots , i+S$, with common coupling constant $\sigma$. Similarly, 
each $y_i-$variable is only coupled to $y_j$-variables, $j=i-S, \cdots , i+S$, with coupling constant $\sigma$, while 
each $\omega_i-$variable is only coupled to $\omega_j$-variables, $j=i-S,  \cdots , i+S$, with coupling constant $\sigma_{\omega}$.
Without loss of generality, we set a common coupling constant, $\sigma$, to the $x_i$ and $y_i$ variables 
while a different one, $\sigma_{\omega}$, governs the frequency exchanges.
 The coupled system of equations takes the form:
\begin{subequations}
\begin{align}
\frac{dx_i}{dt} &= -ax_i+\frac{aR}{\sqrt{x_i^2+y_i^2}}x_i-\omega y_i + \frac{\sigma}{2S}\sum_{j=i-S}^{i+S}\left[ x_j-x_i\right]
\label{eqno04a} \\
\frac{dy_i}{dt} &= -ay_i+\frac{aR}{\sqrt{x_i^2+y_i^2}}y_i+ \omega x_i + \frac{\sigma}{2S}\sum_{j=i-S}^{i+S}\left[ y_j-y_i\right]
\label{eqno04b} \\
\frac{d\omega_i}{dt} &= c (\omega_i -\omega_l)(\omega_i -\omega_c)(\omega_i -\omega_h)+\frac{\sigma_{\omega}}{2S}
\sum_{j=i-S}^{i+S}\left[ \omega_j-\omega_i\right] ,
\label{eqno04c}
\end{align}
\label{eqno04}
\end{subequations}
\noindent where all indices are taken$\mod{N}$. 
All oscillators start from random initial conditions in the $(x,y,\omega)$-variables. Assuming that the $\omega_l$
and $\omega_h$ are the attracting fixed points,
the scenario which is now expected to lead to the formation of chimera states has the following logic:
\begin{figure}[t]
\includegraphics[clip,width=0.85\linewidth,angle=0]{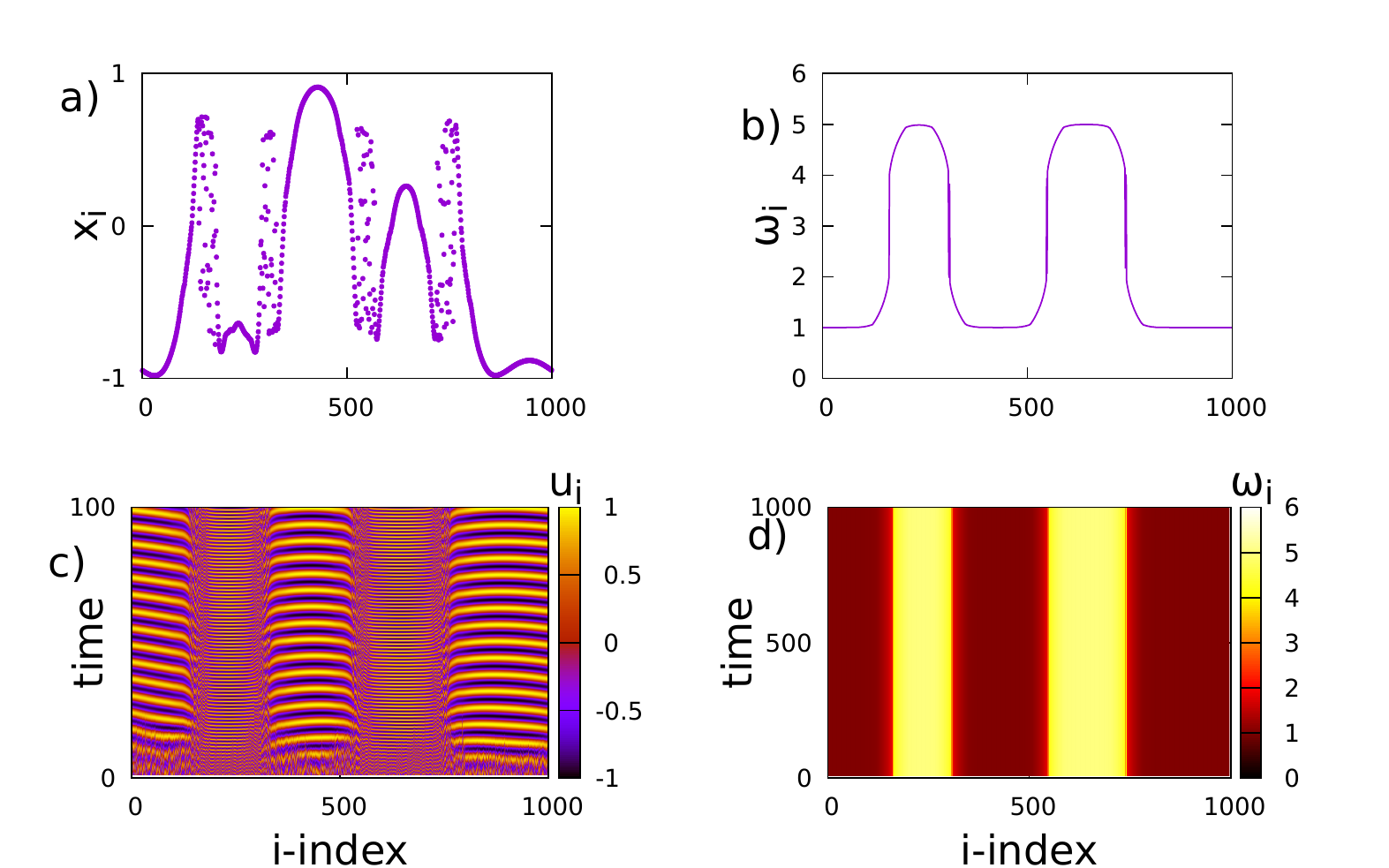}
\caption{\label{fig02} (Color online)
The chimera state formed by the two frequency model: a) $x-$variable profile, b) the $\omega$-variable profile  c)
the space time plot of the $x-$variable and d) the spacetime plot of the $\omega -$variable.
Parameters are: $c=-1$, $\omega_l=1$, $\omega_c=3$, $\omega_h=5$, $R=1$, $S=40$, $a=1.0$, $\sigma =0.5$ 
and $\sigma_{\omega}=3.0$. All simulations start from random initial conditions.
}
\end{figure}

\begin{enumerate}
\item As the system integrates, the frequency $\omega_i$ of each oscillator will fall on one or the other attracting fixed points
 $\omega_l$ or $\omega_h$, 
depending on their initial $\omega_i(t=0)$.
\item Because of the coupling $\sigma_{\omega}$ nearby oscillators will be organized in domains having common frequency,
either $\omega_l$ or $\omega_h$, while the frequencies will alternate in sequential domains.
\item Within each frequency domain the oscillators will have common frequency and due to their coupling, $\sigma$, will
also synchronize in phase ($x$- and $y-$variables).
\item Elements in the borders between two frequency domains will be influenced both by their left and right neighbors
(have different frequencies $\omega_l$ and $\omega_h$) and will oscillate asynchronously, creating the asynchronous domains.
\item Such a chimera state arises, with alternating synchronous domains involving two different frequencies (wells), 
bordered by the asynchronous domains.
\end{enumerate}
\par
As an example, we present in Fig.~\ref{fig02} the chimera state for the working parameter set with specific parameter
values:  $S=40$,  $\sigma =0.5$ and $\sigma_{\omega}=3.0$. In panel a) we present the
$x-$variable profile, in panel b) the $\omega$-variable profile and in (c) the spacetime plot of the
$x-$variable and in panel d) the space time plot of the $\omega$ profile. A chimera having 4 coherent and 4
incoherent domains is formed. Two of the coherent domains have frequencies $\omega_l$ and the other two have $\omega_h$, while the incoherent domains form the borders between the coherent ones. This figure will be used as an exemplary
case in sec. \ref{sec:spectra} for the 
comparative spectral analysis of the nodes belonging to the coherent and incoherent regions.

\par
The scenario realized above might be at the basis of the chimera states observed in other systems.
 It is possible that the combined effects of the nonlinear terms 
and the coupling in these systems, may induce bistability in their frequencies. If this is the case,
then the two-frequency scenarios apply and thus the chimera states are created.

\par Without loss of generality, in the following sections we use the working parameter set: 
$c=-1$ (to ensure of the existence of one repulsive fixed point, $\omega_c$,
surrounded by two attractive ones $\omega_l$ and $\omega_h$), 
$\omega_l=1$, $\omega_c=3$, $\omega_h=5$, $R=1$, $S=40$, $a=1.0$, $\sigma =0.5$ and $\sigma_{\omega}=3.0$. 
Using these parameter values,  in section~\ref{sec:range}
we vary the coupling range $S$ and in section~\ref{sec:freq} the frequency coupling constant to study the 
chimera variations under changes of these parameters. 

\section{Variations with the coupling range}
\label{sec:range}
In this section we keep all parameters fixed to the working set and we monitor the chimera properties
with variation on the coupling range $S$. 

\begin{figure}[H]
\centering
\includegraphics[clip,width=0.8\linewidth,angle=0]{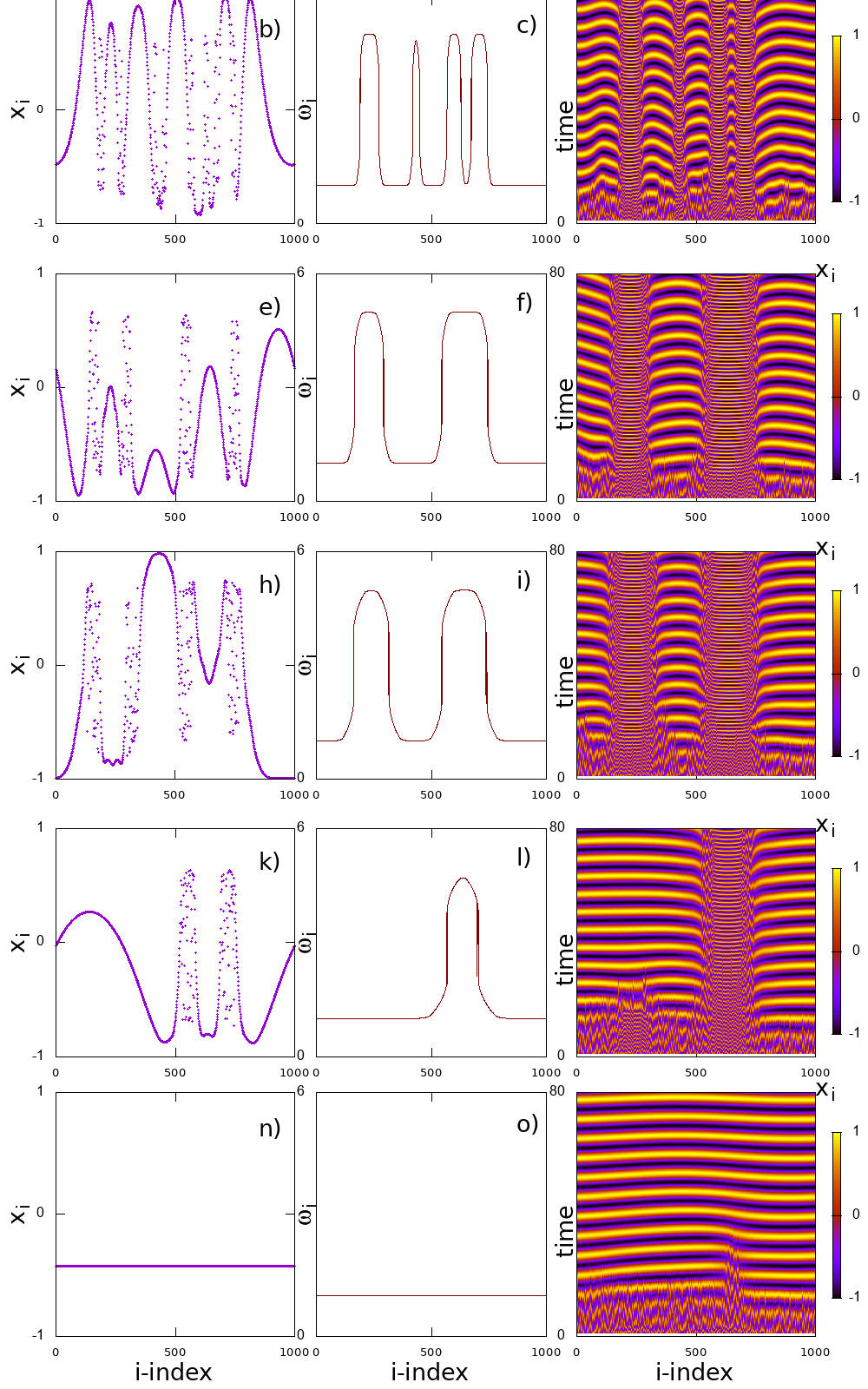}
\caption{\label{fig03} (Color online)
Typical chimera state formed by the two frequency model for different values of the coupling range.
The plots in the left column depict the $x$-variable profiles,
the middle column the mean phase velocities and the right column the spacetime plots of the $x$-variable: 
a) $S=15$, $x-$variable profile, b) $S=15$, the $\omega$-variable profile  c) $S=15$, space time plot of the $x-$variable;
d) $S=30$, $x-$variable profile, e) $S=30$, the $\omega$-variable profile  f) $S=30$, space time plot of the $x-$variable;
g) $S=50$, $x-$variable profile, h) $S=50$, the $\omega$-variable profile  i) $S=50$, space time plot of the $x-$variable;
j) $S=79$, $x-$variable profile, k) $S=79$, the $\omega$-variable profile  l) $S=79$, space time plot of the $x-$variable;
m) $S=90$, $x-$variable profile, n) $S=90$, the $\omega$-variable profile  o) $S=90$, space time plot of the $x-$variable;
Parameters are: $c=-1$, $\omega_l=1$, $\omega_c=3$, $\omega_h=5$, $R=1$, $a=1.0$, $\sigma =0.5$ and $\sigma_{\omega}=3.0$. 
All runs start from random initial conditions.}
\end{figure}

\par  In Fig.~\ref{fig03} we depict the evolution of the chimera states for $S=15$ (top line), $S=30$ (second line),
 $S=50$ (third line), $S=79$ (fourth line), $S=90$ (bottom line). It is clear that small values of the coupling
range $S$ give rise to a large number of coherent (and incoherent) regions, while as we increase the size of $S$
the number of coherent (incoherent) regions decrease. For $S>80$ full synchronization is achieved for the working
parameter set. This is not unexpected. Provided that the system size remains constant,
 for large coupling ranges the exchanges between 
elements cover larger distances, causing communications to larger and larger regions with full synchronization 
as the ultimate state, see Fig.~\ref{fig03}m,n and o.

\par The decrease in the number of coherent (incoherent) domains with increasing $S$ has also been observed in other
systems, such as in FitzHugh-Nagumo \cite{omelchenko:2013}, Leaky Integrate-and-Fire \cite{tsigkri:2017}, the Van der Pol \cite{omelchenko:2015} and other oscillator 
 networks \cite{schoell:2016}. An intuitive understanding of this effect is that the coupling range defines the region where
oscillators interact and thus can synchronize. Therefor,
 the larger the coupling range, the larger the synchronized regions
and consequently fewer synchronous and asynchronous regions can be accommodated by a system of finite, constant 
length. 
\par
\begin{figure}[H]
\includegraphics[clip,width=0.85\linewidth,angle=0]{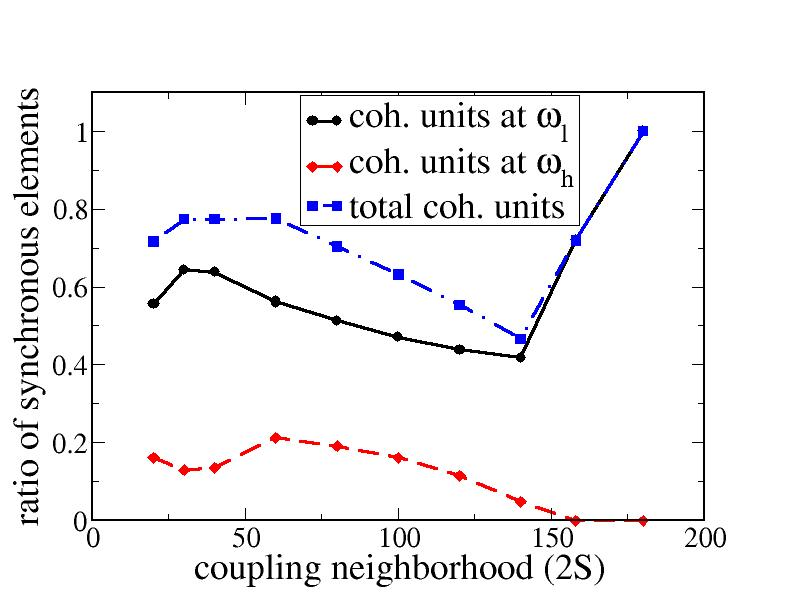}
\caption{\label{fig04} (Color online)
Two frequency chimera properties: The ratio of oscillators having frequency $\omega_l=1$ (solid, black line), 
$\omega_h=5$ (dashed, red line) and the total number of synchronous elements (dashed-dotted, blue line)
as a function of neighborhood size $2S$. Other parameter values as in Fig.~\ref{fig02}.
}
\end{figure}

 In Fig.~\ref{fig04} we present quantitative results on 
the ratios of elements that belong to the lower frequency domain
$\omega_l$ (solid, black line), the higher frequency domain
$\omega_h$ (dashed, red line) and the total number of synchronous elements (dashed-dotted, blue line)
as a function of neighborhood size $2S$. Let us denote by $r=2S/N$, the ratio of coupled elements $2S$
over  the total number of elements $N$ in the network. We may notice three regions where the behavior is distinct:
a) Small sizes of coupling ranges $r<0.03$ (or $2S <30$): Here the number of elements that synchronize
close to $\omega_l$ ($\omega_h$) increases (decreases). b) Intermediate sizes of coupling ranges $0.04<r<0.15$
(or $40<2S<150$): Here the number of elements that synchronize
close to $\omega_l$  decreases and so do the elements that synchronize
close to $\omega_h$. Around $2S=160, \> r=0.16$ all elements that have the highest synchronization frequency disappear
and the system is left with a synchronous domain at $\omega =\omega_l=1$, while the rest of the elements
belong to the asynchronous regime. c) Large coupling ranges
$r>0.16$ or ($2S>160$): Here the number of elements that synchronize at $\omega_l$
gradually increases and reaches full synchronization beyond $r>0.18, \> 2S>180$. This scenario is fairly generic
and for large values of the coupling constant one of the two frequency domains dominates at the expense of the other.

\section{Variations with the coupling constants}
\label{sec:freq}
\par It is interesting to study the variations of this models with the coupling constants. As in most cases of
synchronization in the form of chimera states the coupling constant $\sigma$ which governs the amplitude synchronization
does not affect the chimera multiplicity but only the size of coherent and incoherent domains.
 We shortly discuss
this case using an example in the Appendix \ref{appendixI}. As the exemplary case shows,
 the size of the 
incoherent regions decreases as $\sigma$
increases, leading to full synchronization for large
values of the coupling strength $\sigma$ . 

\begin{figure}[]
\includegraphics[clip,width=0.8\linewidth,angle=0]{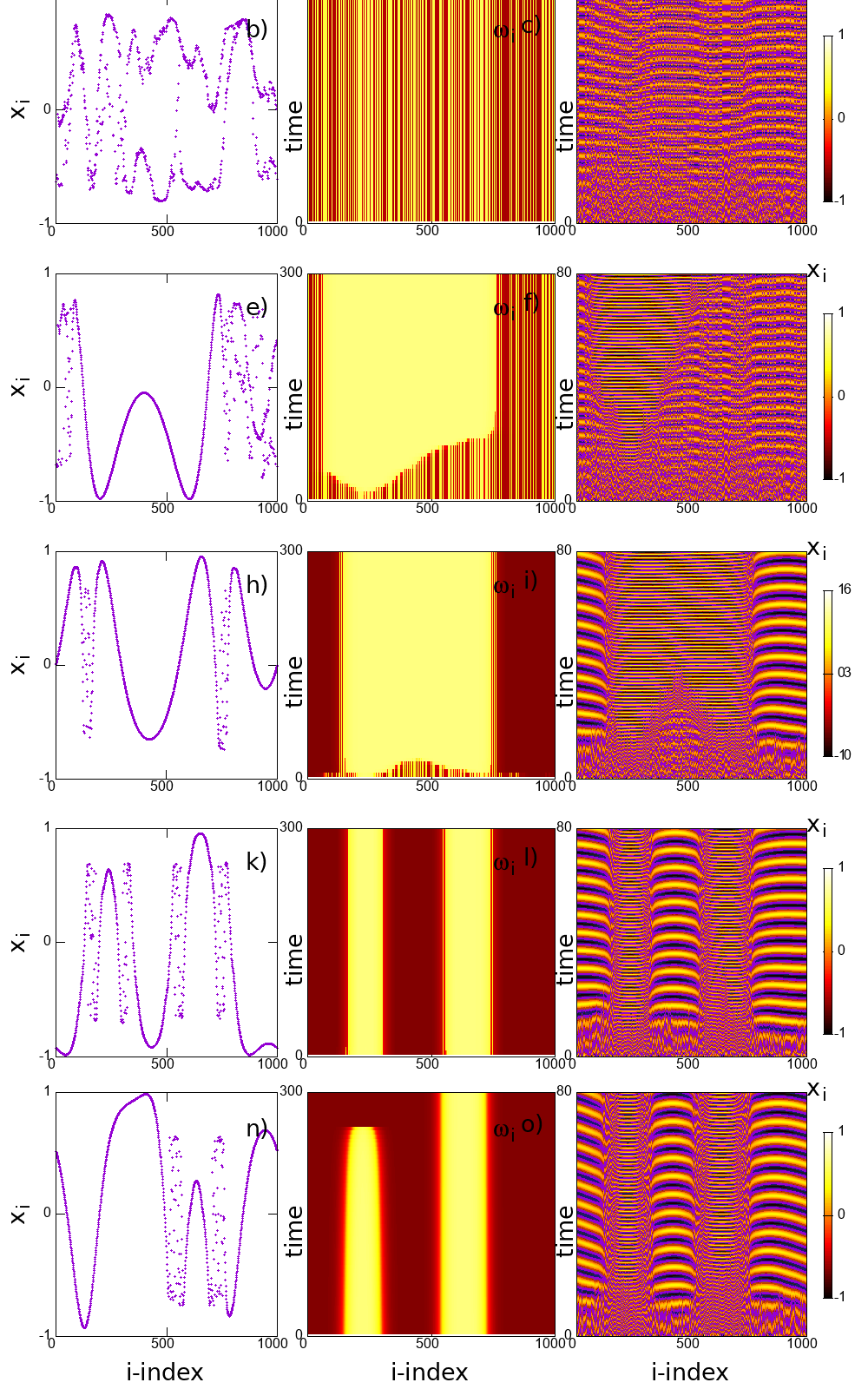}
\caption{\label{fig05} (Color online)
Two frequency chimera properties: Left columns are $x$-profiles, middle columns depict $\omega$-and the right columns depict $x$-space time plots.
a) $\sigma_{\omega}=1.5$, $x-$variable profile, b) $\sigma_{\omega}=1.5$, spacetime plot of $\omega$-variable c) $\sigma_{\omega}=1.5$, space time plot of $x-$variable;
d) $\sigma_{\omega}=1.8$, $x-$variable profile, e) $\sigma_{\omega}=1.8$, spacetime plot of $\omega$-variable,  f) $\sigma_{\omega}=1.8$, space time plot of $x-$variable;
g) $\sigma_{\omega}=2.0$, $x-$variable profile, h) $\sigma_{\omega}=2.0$, spacetime plot of $\omega$-variable,  i) $\sigma_{\omega}=2.0$, space time plot of $x-$variable;
j) $\sigma_{\omega}=2.5$, $x-$variable profile, k) $\sigma_{\omega}=2.5$, spacetime plot of $\omega$-variable,  l) $\sigma_{\omega}=2.5$, space time plot of $x-$variable;
m) $\sigma_{\omega}=5.5$, $x-$variable profile, n) $\sigma_{\omega}=5.5$, spacetime plot of $\omega$-variable, o) $\sigma_{\omega}=5.5$, space time plot of $x-$variable;
for $\sigma =$ a) 0.2, b) 0.6, c) 1.0  and d) 1.5.
Other parameters are: $c=-1$, $\omega_l=1$, $\omega_c=3$, $\omega_h=5$, 
$R=1$, $a=1.0$, $S=40$ and $\sigma =0.5$.
}
\end{figure}

\par After the brief discussion on the amplitude coupling strength, 
we now turn to the more interesting
case of the variations in the frequency coupling range $\sigma_{\omega}$. 
Figure \ref{fig05} provides a first account on the formation of the two frequency
chimera state as we turn on the coupling on the frequency variables. In the absence
or for small values
of the $\sigma_{\omega}$ all oscillators fall fast in their attracting fixed points
($\omega_l$ or $\omega_h$)
and they perform oscillations with these frequencies. Neighboring oscillators are not
affected and the system remains asynchronous in time. Such an example is presented
in the top row of Fig.~\ref{fig05}, with $\sigma_{\omega}=1.5$. 
From the spacetime plot of the mean-phase velocity,
Fig.~\ref{fig05}b, it is evident that all oscillators acquire a constant in time $\omega_i$.
As frequency coupling increases in the second row
of Fig.~\ref{fig05} to $\sigma_{\omega}=1.5$, 
a part of the system develops synchronization in the
highest frequency, $\omega_h =5$, while the rest of the oscillators to the left and
right of the synchronous regions
have mixed frequencies, see Fig.~\ref{fig05}e.  This is
because the
asynchronous regions keep their local frequencies which have been shaped as the
oscillators were attracted by fixed points $\omega_l$ or $\omega_h$. 
The
corresponding $x$-profile, Fig.~\ref{fig05}d, demonstrates coherent motion in the synchronous region and incoherent
outside of it, while the $x$-spacetime plot, Fig.~\ref{fig05}f, indicates that 
even within the incoherent domain small regions of random sizes are formed with
almost coherent temporal behavior. This is not discernible in the $x$-profile but
is visible in the spacetime plot. 
\par By increasing further the frequency coupling to $\sigma_{\omega}=2$,
 see Figs.~\ref{fig05}h,i,g, the
previously asynchronous domain now synchronizes to the higher frequency, $\omega_l$,
thus leading to a chimera state consisting of two coherent domains, one  in the
high frequency, $\omega_h = 5$ and one in the low frequency  $\omega _l=1$. Two
incoherent domains are develop which serve as the borders between the two coherent ones.
The two incoherent domains also serve for continuity purposes to bridge the gap
between the domains where   $\omega_h$ and  $\omega _l$ dominate. As the frequency
coupling increases further  to $\sigma_{\omega}=2.5$,
 see Figs.~\ref{fig05}j,k,l, the high frequency domain splits giving rise to four,
coherent domains bordered by four incoherent ones. By further increasing $\sigma_{\omega}=5.5$
the exchange of $\omega$ variables become dominant and as time increases the
neighboring domains compete and the
lower frequency domains expand in expense of the higher ones, see Figs.~\ref{fig05}j,k,l.
All runs in Fig.~\ref{fig05} start from the same random initial state.

\section{Frequency spectra of coherent and incoherent elements}
\label{sec:spectra}
To investigate the transition from the coherent to incoherent regions and to check which
precise frequencies are present in each region we investigate the Fourier spectra of
different oscillators belonging to the core of the coherent and
others occupying borderline regions.
\par We analyze here the results in Fig.~\ref{fig02} and plot the Fourier spectra of nodes $i=440, \> 650$
and $740$. The first one, $i=440$, is centrally located in the coherent domain which oscillates with $\omega =1$,
the second, $i=650$, belongs to the coherent domain with $\omega =5$, while the third one, $i=740$, occupies a
position in between the two, in the incoherent region which serves as a transition region 
between the two domains. The results are shown in Fig.~\ref{fig06}.

\par The left panel of Fig.~\ref{fig06} depicts the Fourier amplitude of node $i=440$. Only one peak is clearly
seen at frequency values $f=\omega /2\pi =0.167402$, or $\omega \sim 1$, as expected for the oscillators which
have fallen in the basin of attraction of the lowest frequency, $\omega_l =1$. The middle panel, Fig.~\ref{fig06}b, depicts the Fourier amplitude of node $i=650$. Also here, only one peak is developed
 at frequency values $f=\omega /2\pi =0.789951$, or $\omega \sim 5$, as expected for the oscillators which
have fallen in the basin of attraction of the lowest frequency, $\omega_h =5$. The right panel, Fig.~\ref{fig06}c, depicts the Fourier amplitude of node $i=740$. Here four peaks appear at
 frequency values i) $f=\omega /2\pi =0.167402$ ($\omega \sim 1$), ii) $f=\omega /2\pi =0.79$ ($\omega \sim 5$),
iii) $f=\omega /2\pi = 0.642$ ($\omega \sim 4.03$) and iv) $f=\omega /2\pi =0.496$ ($\omega \sim 3.1$).
In these regions the oscillators are developing mixed behavior and present mixed oscillatory
characteristics, drawing both from the low frequency dynamics (Case i) and the high frequency dynamics (Case ii).
In addition two more peaks with high amplitude are developed which 
can be considered as linear combinations of the $\omega_h$ and $\omega_l$ values. E.g., the peak with the highest
amplitude in the right panel of Fig.~{\ref{fig06} which corresponds to $\omega =4.03$ is approximately equal
to $(\omega_l+\omega_h )/2$. Within the
incoherent region, the closer the element is to the high (low) frequency domain, the higher the amplitude of the 
 corresponding peak is (images not shown). 
\par These findings corroborate the intuitive argument
 that the two different types of coherent domains are formed due to the bistability of
the oscillator frequency caused by the addition of the linear coupling terms to the nonlinear dynamics. In this
view, the incoherent regions serve for the purpose of continuity when passing from the lower to the higher frequency
domains (and the opposite).  That is the reason why they have a gradient in frequencies, giving rise to the asynchronous
incoherent domains in the  $x$-variable profile (and $y-$variable profile, not shown).  

\par At this point one would argue that a number of chimera states do not present two-level synchronization,
but they demonstrate an arc-shaped mean phase velocity profile. Given the present results, 
it is not possible to conclude
 whether the  arc-shape $\omega$-profile of the chimera states in the Kuramoto \cite{kuramoto:2002}
or the FitzHugh Nagumo \cite{omelchenko:2013} models emerges as
an incompletely formed higher (or lower) frequency domain, or if some other phenomenon is responsible for this
profile.
\medskip
\medskip

\begin{figure}[]
\includegraphics[clip,width=0.32\linewidth,angle=0]{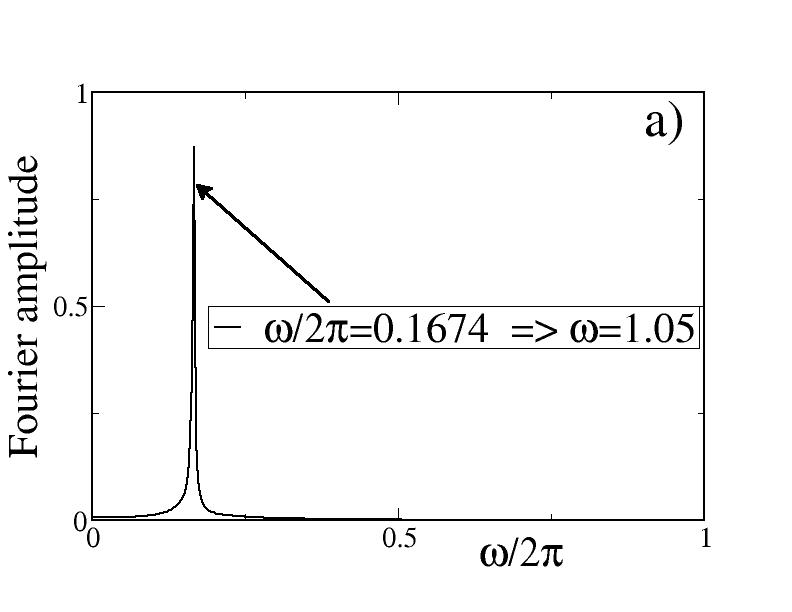}
\includegraphics[clip,width=0.32\linewidth,angle=0]{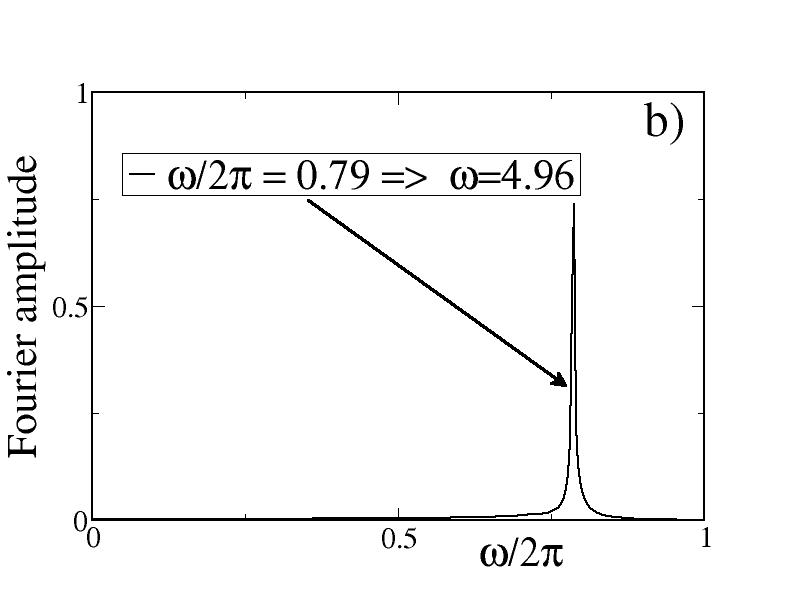}
\includegraphics[clip,width=0.32\linewidth,angle=0]{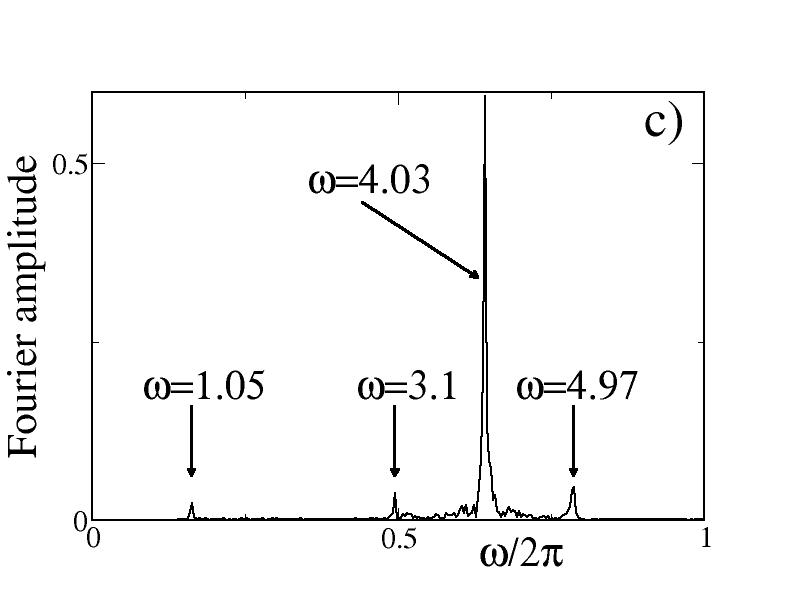}
\caption{\label{fig06} (Color online)
Spectral analyses of three nodes referring to the simulation reported
in Fig.~\ref{fig02}: a) element $i=440$
belonging to the coherent region of low frequency, b)  element $i=650$
belonging to the coherent region of high frequency and c)  element $i=740$
belonging to incoherent region. All parameter values refer to Fig.~\ref{fig02}.
}
\end{figure}

\par As an additional indicator of the different frequencies dominating in the coherent regions, 
we plot in Fig.~\ref{fig07} the Fourier amplitude of the peak which corresponds
to the low mean
phase velocity,
$\omega_l \sim 1$ (black line) and  to the high one, $\omega_h \sim 5$ (red line).
Comparing Figs.~\ref{fig07} and ~\ref{fig02}b we note that the Fourier amplitude of the peak at frequencies
$\omega \sim 1$  dominates in the coherent regions of low $\omega$ in Fig.~\ref{fig02}b
and vanishes gradually in domains of high frequency (see Fig.~\ref{fig07}, black line). Similarly,
the red line which depicts the Fourier amplitude extracted from the the peak at frequencies
$\omega \sim 5$
reaches maximum values in the coherent domains of high $\omega$ in Fig.~\ref{fig02}b and vanishes in the
domains of low frequency. In the intermediate domains, between the low and high frequency ones, the Fourier
amplitudes acquire intermediate values. At the same time new peaks appear, as shown in Fig.~\ref{fig06}c,
whose spectra are the linear combinations of the $\omega_h$ and $\omega_l$ values. These additional
spectral lines, clearly appearing in ~\ref{fig06}c, are omitted in Fig.~\ref{fig07} for clarity reasons.
\par The results in this section indicate that the calculation of the
Fourier spectra is a laborious but reliable method to identify 
qualitatively and qualitatively the different coherent and incoherent domains
and to assert the presence (or absence) of a chimera state.

\begin{figure}[]
\includegraphics[clip,width=0.45\linewidth,angle=0]{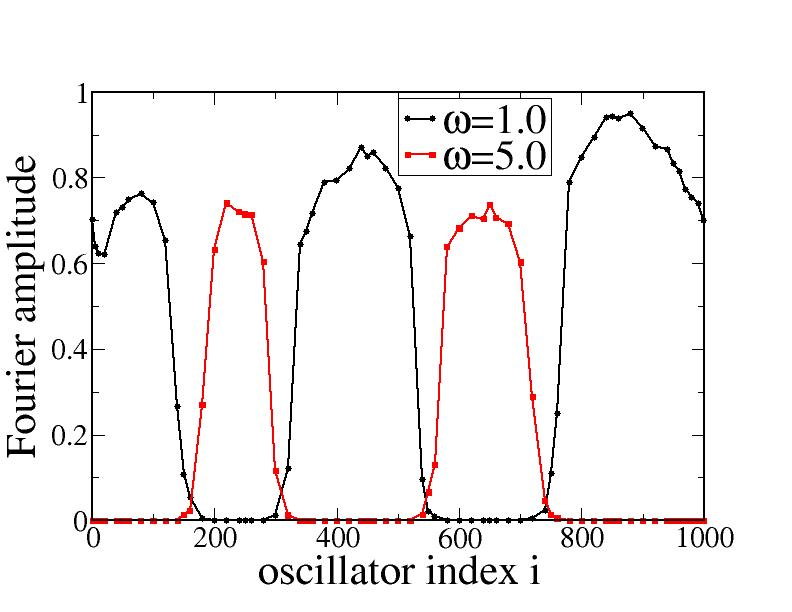}
\caption{\label{fig07} (Color online)
Fourier amplitudes of the dynamics of all elements for the numerical results reported in Fig.~\ref{fig02}.
 The black line depicts the Fourier amplitude of the peak which corresponds to the low mean
phase velocity,
$\omega_l \sim 1$ and the red line to the high one, $\omega_h \sim 5$. 
All parameter values refer to Fig.~\ref{fig02}.
}
\end{figure}

\section{Conclusions}
\label{sec:conclusions}

In the present study we first introduced a model oscillator whose trajectory tends 
exponentially to a closed limiting orbit
with well defined frequency and radius. As numerical results indicate,
no chimera states arise when such oscillators are coupled, but the system tends to either full synchronization
or to full disorder. When a third equation is added which allows the oscillators to choose between two different
frequencies, a higher and a lower ones, then domains are formed where the high and low frequencies dominate.
The domains of different frequencies are mediated by incoherent domains where the oscillator frequencies gradually
increase or decrease to bridge the frequency gap between adjacent regions and to maintain continuity in the system.
This scenario can serve as a general mechanism for formation of chimera states. Even in cases where this mechanism is
not explicit, the introduction of coupling together with the nonlinearity in the dynamics may induce bistability 
in the oscillator frequency creating, in this way, indirectly, two-level synchronization 
and corresponding chimera states.
\par The present model can be easily extended to form multi-leveled chimeras, by allowing the oscillators
to occupy many different frequency levels.
%to occupy/equilibrate into/attain??
\par Finally, in Ref.~\cite{provata:2018} except for the case of the exponential relaxation to the limiting orbit, power law relaxation has been
introduced. Power laws take much longer (infinite time) to reach the final
oscillatory trajectory. It would be interesting to investigate
whether this power law relaxation
dynamics leads also to the formation of chimera states and, further-on, if frequency multistability 
can lead to multi-leveled chimera states. 

\section*{Acknowledgements}

\par The author would like to thank Dr. J. Hizanidis for helpful discussions.
This work was supported in part by
the project MIS 5002772, implemented under the Action ``Reinforcement of
the Research and Innovation Infrastructure'', funded by the Operational Programme
 ``Competitiveness, Entrepreneurship and Innovation'' (NSRF 2014-2020) 
and co-financed by Greece and the European Union (European Regional Development Fund).
This work was supported by computational time granted from the Greek Research \& Technology Network (GRNET) 
in the National HPC facility - ARIS - under project CoBrain4 (project ID: PR007011).

%\bibliography{./provata.bib}

\appendix
\section{The role of the amplitude coupling constant}
\label{appendixI}
\renewcommand{\thefigure}{A\arabic{figure}}
\setcounter{figure}{0}
\par  As in most cases of local 
synchronization in the form of chimera states, the coupling constant $\sigma$ which governs the amplitude synchronization
does not affect the chimera multiplicity but only the size of coherent and incoherent domains. As an exemplary case we present here simulation results using the following parameter set:
$c=-1$, $\omega_l=1$, $\omega_c=3$, $\omega_h=5$, $R=1$, $a=1.0$, $S=40$ and $\sigma_{\omega}=3.0$,
with variable size of $\sigma =0.2, \> 0.6, \> 1.0 $ and $1.5$.
In Fig.~\ref{fig-append} the spacetime plots of the variable $x$
are presented in the four cases. For small $\sigma$ values the incoherent regions extend to a large
number of oscillator, while the size of the incoherent domains
decreases for larger $\sigma$-values. 
\begin{figure}[h]
\includegraphics[clip,width=0.85\linewidth,angle=0]{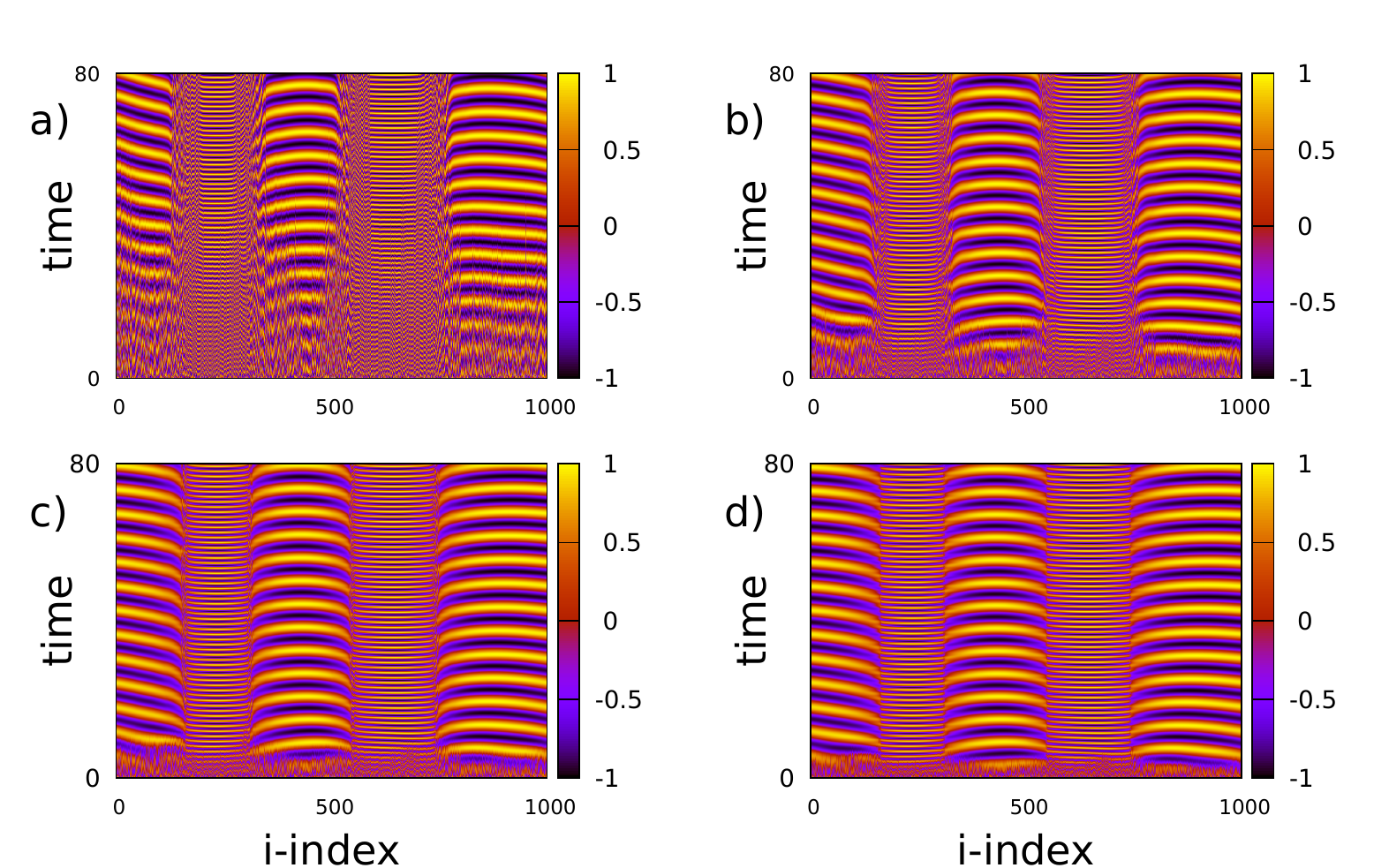}
\caption{\label{fig-append} (Color online)
Two frequency chimera properties: Spacetime plots for $\sigma =$ a) 0.2, b) 0.6, c) 1.0  and d) 1.5.
Other parameters are: $c=-1$, $\omega_l=1$, $\omega_c=3$, $\omega_h=5$, 
$R=1$, $a=1.0$, $S=40$ and  $\sigma_{\omega}=3.0$.
}
\end{figure}

\end{document}